# Superconductivity in the high-entropy ceramics $Ti_{0.2}Zr_{0.2}Nb_{0.2}Mo_{0.2}Ta_{0.2}C_x$ with possible nontrivial band topology

*Lingyong Zeng[1,#], Xunwu Hu[2,#], Yazhou Zhou[7,#], Mebrouka Boubeche[3], Ruixin Guo[4,5], Yang Liu[6], Si-Chun Luo[6], Shu Guo[4,5], Kuan Li[1], Peifeng Yu[1], Chao Zhang[1], Wei-Ming Guo[6], Liling Sun[7,*], Dao-Xin Yao[2,5*], Huixia Luo[1,*]*

([#] These authors contributed equally to this work.)

L. Zeng, K. Li, P. Yu, C. Zhang, H. Luo
School of Materials Science and Engineering, State Key Laboratory of Optoelectronic Materials and Technologies, Key Lab of Polymer Composite & Functional Materials, Sun Yat-Sen University, No. 135, Xingang Xi Road, Guangzhou, 510275, P. R. China
E-mail: luohx7@mail.sysu.edu.cn

X. Hu, D.-X. Yao
Guangdong Provincial Key Laboratory of Magnetoelectric Physics and Devices, Center for Neutron Science and Technology, School of Physics, Sun Yat-Sen University, Guangzhou, 510275, China
E-mail: yaodaox@mail.sysu.edu.cn

M. Boubeche
Songshan Lake Materials Laboratory, Building A1, University Innovation Town, Dongguan City, Guang Dong Province, 523808 China

R. Guo, S. Guo
Shenzhen Institute for Quantum Science and Engineering, Southern University of Science and Technology, Shenzhen 518055, China;
International Quantum Academy, Shenzhen 518048, China

Y. Liu, S.-C. Luo, W.-M. Guo
School of Electromechanical Engineering, Guangdong University of Technology, Guangzhou 510006, China

Y. Zhou, L. Sun
Institute of Physics, Chinese Academy of Sciences, Beijing 100190, China
E-mail: llsun@iphy.ac.cn

**Abstract:** Topological superconductors have drawn significant interest from the scientific community due to the accompanying Majorana fermions. Here, we report the discovery of electronic structure and superconductivity in high-entropy ceramics $Ti_{0.2}Zr_{0.2}Nb_{0.2}Mo_{0.2}Ta_{0.2}C_x$ ($x$ = 1 and 0.8) combined with experiments and first-principles calculations. The $Ti_{0.2}Zr_{0.2}Nb_{0.2}Mo_{0.2}Ta_{0.2}C_x$ high-entropy ceramics show bulk type-II superconductivity with $T_c$ about 4.00 K ($x$ = 1) and 2.65 K ($x$ = 0.8), respectively. The specific heat jump ($\Delta C/\gamma T_c$) is equal to 1.45 ($x$ = 1) and 1.52 ($x$ = 0.8), close to the expected value of 1.43 for the BCS superconductor in the weak coupling limit. The high-pressure resistance measurements show that a robust superconductivity against high physical pressure in $Ti_{0.2}Zr_{0.2}Nb_{0.2}Mo_{0.2}Ta_{0.2}C$, with a slight $T_c$ variation of 0.3 K within 82.5 GPa. Furthermore, the first-principles calculations indicate that the Dirac-like point exists in the electronic band structures of $Ti_{0.2}Zr_{0.2}Nb_{0.2}Mo_{0.2}Ta_{0.2}C$, which is potentially a topological superconductor. The Dirac-like point is mainly contributed by the *d* orbitals of transition metals M and the *p* orbitals of C. The high-entropy ceramics provide an excellent platform for the fabrication of novel quantum devices, and our study may spark significant future physics investigations in this intriguing material.

## 1. Introduction

It is always keening for condensed matter scientists to discover new materials and explore their unique physical properties. Superconductivity (SC), in combination with topology, is expected to exhibit new types of quasiparticles, such as non-Abelian Majorana zero modes or fractional charge and spin currents.[1-4] The experimental realization of topological SC will provide an excellent platform for developing fault-tolerant quantum computing techniques.[5]

However, searching for topological superconductors (TSCs) has been challenging. The ways towards realizing topological SC have been adopted: finding SC with nontrivial intrinsic topology or combining the conventional SC with other nontrivial topological band structures (e.g., $Bi_2Se_3/NbSe_2$, $Bi_2Te_3/NbSe_2$)[6,7] or pressurizing/doping topological materials (topological insulator, topological Weyl semimetal, topological Dirac semimetal).[8-16] The existence of nontrivial topology in intrinsic superconducting materials offers the possibility to realize TSCs, preventing the complexity of fabricating a proximity-coupled heterostructure of a superconductor and topological insulator. There have been observations of Majorana zero modes in iron-based TSCs ($FeTe_{1-x}Se_x$, $CaKFe_4As_4$, and LiFeAs).[17-22] In addition, both predicted and experimental intrinsic TSCs are exceedingly rare. Most of them can only achieve SC or suitable topological surface states near the Fermi energy ($E_F$) by doing. It is highly urgent to search for more intrinsic TSC candidates with high superconducting critical temperature ($T_c$) and topological surface states near $E_F$. In recent, some strong candidates for TSC have been found in compounds formed by the IVA group elements and metals, such as $AuSn_4$,[23,24] $Au_2Pb$,[25] $PtPb_4$,[26] $Ta_3$(Sn, Pb),[27] $BaSn_5$,[28] (Ta, Nb)RuSi,[29] and binary transition-metal carbides (TMCs).[30-34] Among these TSCs, TMCs have a relatively high $T_c$. The type-II Dirac semimetal states were proposed to exist in the band structure of NbC and TaC, which are well-known comparable high $T_c$ ~ 11.5 K and 10.6 K superconductors.[30,32,34] The first-principles calculations also indicate that *s*-wave Bardeen-Cooper-Schrieffer (BCS) SC with $T_c$ ~ 14 K and nontrivial band topology coexist in cubic *a*-MoC.[33]

The high-entropy alloy (HEA) concept was developed in 2004 [35], and since then, an entropy stabilization concept has been used to prepare high-entropy ceramics (HECs) as well.[36,37] HECs are the solid solution of five or more cationic or anionic sublattices with a high configuration entropy.[36,37] High-entropy transition metal carbide ceramics (HECCs), as one of the most intensively researched subsets of these materials, generally exhibit superior mechanical and physical properties, such as high hardness, low thermal conductivity, excellent elevated-temperature flexural strength, and good resistance to high-temperature oxidation and wear.[36-40] These high-entropy materials, a form of multi-carbide solid solution, have drawn widespread attention recently due to their vast potential and broad industrial application prospects. However, the intense research on HECCs has primarily focused on their mechanical properties. The physical properties of HECs, especially SC and topological properties, are still worth exploring.

In this study, HECs of $Ti_{0.2}Zr_{0.2}Nb_{0.2}Mo_{0.2}Ta_{0.2}C_x$ ($x$ = 1 and 0.8) with a single-phase NaCl-type structure were prepared by a spark plasma sintering method. We report our discovery and investigation of the HEC superconductors $Ti_{0.2}Zr_{0.2}Nb_{0.2}Mo_{0.2}Ta_{0.2}C_x$ ($x$ = 1 and 0.8), which shows bulk type-II SC with $T_c$ about 4.00 K ($x$ = 1) and 2.65 K ($x$ = 0.8), respectively. Considering the extraordinary properties of the HECCs mentioned above, the discovery of SC and topological band structures in these HECCs would make them an excellent platform for novel quantum device fabrication.

## 2. Experimental Section

The $Ti_{0.2}Zr_{0.2}Nb_{0.2}Ta_{0.2}Mo_{0.2}C_x$ starting powders were synthesized via the carbothermal reduction method,[52] utilizing molar ratio of $MoO_3$, $Ta_2O_5$, $Nb_2O_5$, $ZrO_2$, $TiO_2$, and graphite powders as the precursor materials. These molar ratios used was based on the equation as follows: $TiO_2 + ZrO_2 + 0.5Nb_2O_5 + 0.5Ta_2O_5 + MoO_3 + (5x+12)C \rightarrow 5\ Ti_{0.2}Zr_{0.2}Nb_{0.2}Ta_{0.2}Mo_{0.2}C_x + 12CO$ (g) (where $x$ = 1 or 0.8). The raw powders were subjected to ball milling in anhydrous ethanol for a duration of 24 hours, employing $Si_3N_4$ balls. Subsequently, the powder mixtures were dried using a rotary evaporator and sieved through a 100-mesh sieve. The resulting precursors then were placed in a graphite vacuum furnace and subjected to a temperature of 1650°C for 3h to synthesize $Ti_{0.2}Zr_{0.2}Nb_{0.2}Ta_{0.2}Mo_{0.2}C_x$ powders. A suitable quantity of these powders was then loaded into a graphite mold within the spark plasma sintering (SPS) furnace to sinter $Ti_{0.2}Zr_{0.2}Nb_{0.2}Ta_{0.2}Mo_{0.2}C_x$ samples. The powders were then heated to a temperature of 2000 °C for a duration of 10 min, with a heating and cooling rate of 100 °C/min, under an atmosphere of 1 atm Ar.

PXRD data were taken on the MiniFlex of Rigaku at a scanning rate of 1°/min. Through Rietveld refinements in Fullprof suit software, lattice parameters were obtained. Chemical composition was estimated with SEM-EDX with an electron acceleration voltage of 20 KV. The temperature-dependent electrical resistivity and magnetic susceptibility and heat capacity, were measured by a physical property measurement system (PPMS, Quantum Design. Inc). The resistance measurements were performed with a four-probe method. The magnetization and heat capacity measurements use small pieces of sample. The high-pressure resistance measurements were performed at the high-pressure station equipped with a diamond anvil cell at Synergetic Extreme Condition User Facility. In the measurements, the standard four-probe electrodes (platinum foils) were applied to the samples, and the pressure was determined by the ruby fluorescence method.[53] For all resistivity measurements at ambient pressure, platinum wires were connected to the sample with silver paint.

We perform the calculations using the experimental lattice structure parameters. The Ti/Zr/Nb/Mo/Ta and C atoms are fixed in the observed position 4a (0, 0, 0) and 4b (1/2, 1/2, 1/2), respectively. The chemically disordered solutions of $Ti_{0.2}Zr_{0.2}Nb_{0.2}Hf_{0.2}Ta_{0.2}C_x$ HECs are modeled by the “mcsqs” code of the Alloy Theoretic Automated Toolkit (ATAT).[54] The 2×2×2 supercell with 64 atoms was adopted. The electronic structure properties calculations are performed using the Vienna *ab initio* simulation package (VASP) code[55,56] based on density functional theory (DFT). For exchange-correlation functions, the generalized gradient approximation (GGA) in the form of Perdew-Burke-Ernzerhof (PBE)[57] is adopted. The projector augmented-wave (PAW) method[58] with a 400 eV plane-wave cutoff energy is employed. For Brillouin zone sampling, a Γ-centered 5 × 5 × 5 k-points mesh within the Monkhorst-Pack scheme is used in the self-consistent process. Convergence criteria for the electronic self-consistent iteration are set to $10^{-6}$ eV. Spin–orbit coupling (SOC) is used in the calculations of electronic band structure properties.

## 3. Results and Discussion

**Figure 1**a exhibits the powder X-ray diffraction (PXRD) data of the $Ti_{0.2}Zr_{0.2}Nb_{0.2}Mo_{0.2}Ta_{0.2}C_x$ samples. All the diffraction peaks of $Ti_{0.2}Zr_{0.2}Nb_{0.2}Mo_{0.2}Ta_{0.2}C_x$ ($x$ = 1 and 0.8) are indexed on the space group $Fm\bar{3}m$. A decrease in the $x$ concentration causes the peak position of (111) to shift towards the lower angle side. Figure 1b shows the Rietveld refinement profile of the $Ti_{0.2}Zr_{0.2}Nb_{0.2}Mo_{0.2}Ta_{0.2}C$ sample, which displays reliable, refined results with $\chi^2$= 2.84, $R_{wp}$ = 3.49 %, and $R_p$ = 2.67 %. The lattice parameters are a = 4.4573(3) Å for $x$ = 1 and a = 4.4479(8) Å for $x$ = 0.8, respectively. The inset of Figure 1b displays the simplified schematic diagram of the NaCl-type structure for the $Ti_{0.2}Zr_{0.2}Nb_{0.2}Mo_{0.2}Ta_{0.2}C$ sample. The carbon element occupies the anion position, while five metal elements likely share a cation position. We further carried out the scanning electron microscopy (SEM) and energy-dispersive X-ray spectroscopy (EDX) characterization of $Ti_{0.2}Zr_{0.2}Nb_{0.2}Mo_{0.2}Ta_{0.2}C_x$ ($x$ = 1 and 0.8) fresh cross-section to check the homogeneity and actual ratio of the compounds. As seen in the Supporting information **Figure S**1 and **Figure S**2, all constituent elements are homogeneously distributed. The proportion of each metal element is close (See **Figure S**3). Note that this method cannot accurately determine its content since carbon has a light mass and is most likely a contaminant in the EDX analysis. Nevertheless, the EDX results showed

that the carbon content of the $x$ = 1 sample is higher than the carbon content of the $x$ = 0.8 sample. **Figure 1c-d** shows the temperature dependencies of resistivity for $Ti_{0.2}Zr_{0.2}Nb_{0.2}Mo_{0.2}Ta_{0.2}C_x$ ($x$ = 1 and 0.8) samples. A sharp resistivity drop is observed in both cases, indicating the superconducting transition. The zero-resistivity was achieved at 4.00 K for $x$ = 1 and 2.68 K for $x$ = 0.8. The normal resistivity decreases only slightly with a near temperature independent, similar to that observed in HEA superconductors.[41,42] The residual resistivity ratio (RRR) value for $Ti_{0.2}Zr_{0.2}Nb_{0.2}Mo_{0.2}Ta_{0.2}C_x$ ($x$ = 1 and 0.8) samples is close to 1.

The temperature-dependent magnetic susceptibility was measured under 20 Oe in the zero-filed cooling (ZFC) for $x$ = 1 (**Figure 2**a) and $x$ = 0.8 (**Figure 2**b). To get a more accurate value of the superconducting shielding fraction, the demagnetization factors (N) for $Ti_{0.2}Zr_{0.2}Nb_{0.2}Mo_{0.2}Ta_{0.2}C_x$ samples are estimated to be 0.72 ($x$ = 1) and 0.47 ($x$ = 0.8) respectively, by using N = 1 + 1/(4πs), where s is the slope of linear fitting in the field-dependent volume magnetization curve at 1.8 K. We also calculate the theoretical N value using the equation $N^{-1} = 1 + \frac{3}{4}\frac{c}{a}(1 + \frac{a}{b})$,[43] where 2a × 2b × 2c is the geometric parameters of the cuboid sample. The theoretical N values are calculated to be 0.67 ($x$ = 1) and 0.43 ($x$ = 0.8), respectively, consistent with the actual values. The resulting diamagnetic signal with a clear transition to a superconducting state is close to 100 % Meissner volume fraction, indicating the bulk nature of SC in $Ti_{0.2}Zr_{0.2}Nb_{0.2}Mo_{0.2}Ta_{0.2}C_x$ samples. The onset diamagnetic transition temperatures are 4.00 K for $x$ = 1 and 2.65 K for $x$ = 0.8, which agree well with that from resistivity data.

**Figure 2c-d** shows isothermal magnetization curves over different temperatures below the $T_c$ for $Ti_{0.2}Zr_{0.2}Nb_{0.2}Mo_{0.2}Ta_{0.2}C_x$ ($x$ = 1 and 0.8) samples. The lower critical

fields ($\mu_0H_{c1}(0)$) are obtained from the fields where the M(H) deviates from the linearly field-dependent behavior (Meissner line), i.e., the magnetic flux starts to penetrate the SC body. All the uncorrected lower critical fields, $\mu_0H_{c1}$*, with the corresponding temperatures, are plotted in **Figure 2**e for $x$ = 1 and Figure 2f for $x$ = 0.8. The data points are modeled with the GL relation: $\mu_0 H_{c1}^*(T) = \mu_0 H_{c1}^*(0)(1-(T/T_c)^2)$, giving $\mu_0H_{c1}$*(0) = 8.1(1) mT for $x$ = 1 and $\mu_0H_{c1}$*(0) = 5.3(1) mT for $x$ = 0.8. Considering the demagnetization factor, the real $\mu_0H_{c1}(0)$ can be deduced from the $\mu_0H_{c1}$* with the formula $\mu_0H_{c1}(0) = \mu_0H_{c1}^*(0)/(1-N)$. The estimated $\mu_0H_{c1}(0)$ = 28.9(6) mT for $x$ = 1 and $\mu_0H_{c1}(0)$ = 10.0(2) mT for $x$ = 0.8.

The low-temperature resistivity under different magnetic fields for $Ti_{0.2}Zr_{0.2}Nb_{0.2}Mo_{0.2}Ta_{0.2}C_x$ ($x$ = 1 and 0.8) samples is presented in **Figure 3**a and Figure 3b, respectively. Upon applying the magnetic field, the $T_c$ decreases steadily for both HECs. **Figure 3**c-d shows the upper critical fields $\mu_0H_{c2}$ plotted as a function of the estimated $T_c$ values for $Ti_{0.2}Zr_{0.2}Nb_{0.2}Mo_{0.2}Ta_{0.2}C_x$ ($x$ = 1 and 0.8) samples. The zero temperature for a type-II superconductor in dirty limit can be calculated with the Werthamer Helfand Hohenberg (WHH) theory: $\mu_0H_{c2}(0) = -0.693T_c(\frac{d\mu_0 H_{c2}}{dT})|_{T = T_c}$. The extrapolated slopes near $T_c$ are $\frac{d\mu_0 H_{c2}}{dT}$ = -0.93(3) T/K for $x$ = 1, and $\frac{d\mu_0 H_{c2}}{dT}$ = -0.92(7) T/K for $x$ = 0.8. Thus, based on the slope and $T_c$, we have $\mu_0H_{c2}(0)$ = 2.5(9) T for $x$ = 1 and 1.7(0) T for $x$ = 0.8. The $\mu_0H_{c2}(0)$ is also deduced by extrapolating the data based on the GL model: $\mu_0 H_{c2}(T) = \mu_0 H_{c2}(0) * \frac{1-(T/T_c)^2}{1+(T/T_c)^2}$, giving 3.2(6) T and 2.3(4) T for $x$ = 1 and $x$ = 0.8, respectively. According to the equation $\mu_0H^P = 1.85*T_c$, the Pauli paramagnetic limits are 7.40 T and 4.90 T for $x$ = 1 and $x$ = 0.8, respectively.

Various superconducting parameters can be calculated using $\mu_0H_{c1}(0)$ and $\mu_0H_{c2}(0)$. First, according to the formula $\xi_{GL}^2(0) = \frac{\Phi_0}{2\pi\mu_0 H_{c2}(0)}$, where $\Phi_0$ = h/2e represents the flux quantum, the GL coherence length ($\xi_{GL}(0)$) is determined to be 100.5(3) Å and 118.6(5) Å for $x$ = 1 and $x$ = 0.8, respectively. Second, the GL penetration depth at zero K, $\lambda_{GL}(0)$, is obtained 1186 Å and 2192 Å for $x$ = 1 and $x$ = 0.8, respectively, using the expression: $\mu_0 H_{c1}(0) = \frac{\Phi_0}{4\pi\lambda_{GL}^2(0)} ln \frac{\lambda_{GL}(0)}{\xi_{GL}(0)}$. Third, the GL parameter, $K_{GL}(0) = \frac{\lambda_{GL}(0)}{\xi_{GL}(0)}$, can be estimated to be 11.8 ($x$ = 1) and 18.5 ($x$ = 0.8), which is larger than $1/\sqrt{2}$, suggesting that $Ti_{0.2}Zr_{0.2}Nb_{0.2}Mo_{0.2}Ta_{0.2}C_x$ HECs are strongly type-II superconductors. Table S1 summarizes all the gathered normal and superconducting parameters for $Ti_{0.2}Zr_{0.2}Nb_{0.2}Mo_{0.2}Ta_{0.2}C_x$ samples. For comparison, the relevant superconducting parameters of previously reported HECCs are also listed in Table S1.[44,45]

The low-temperature specific heat measurements under applied magnetic fields of 0 and 5 T were performed to confirm the bulk nature of the SC. The obvious anomaly in the 0 T heat capacity (**Figure 4**a and Figure 4b), corresponding with the emergence of the superconducting state, can be observed in $Ti_{0.2}Zr_{0.2}Nb_{0.2}Mo_{0.2}Ta_{0.2}C_x$ ($x$ = 1 and 0.8) samples. Based on the equal entropy construction, we find that the $T_c$ = 3.98 K for $x$ = 1 sample and $T_c$ = 2.49 K for $x$ = 0.8 sample. The heat capacity data is fitted well with the Debye model, $C_p/T = \gamma + \beta T^2 + \eta T^4$, where two-term, $\beta T^2 + \eta T^4$, are used to express the phonon contribution, and $\gamma$ is the normal state electronic specific heat coefficient. The best fits give $\gamma$ = 2.471(5) mJ·mol$^{-1}$·K$^{-2}$, $\beta$ = 0.010(6) mJ·mol$^{-1}$·K$^{-4}$ for $x$ = 1 sample, and $\gamma$ = 2.206(2) mJ·mol$^{-1}$·K$^{-2}$, $\beta$ = 0.012(9) mJ·mol$^{-1}$·K$^{-4}$ for $x$ = 0.8 sample. Then, another important superconducting parameter, specific heat jump ($\Delta C/\gamma T_c$) at $T_c$, can be determined. The $\Delta C/\gamma T_c$ is equal to 1.45 ($x$ = 1) and 1.35 ($x$ = 0.8), close to the expected value of 1.43 for the BCS superconductor, verifying the bulk nature of the SC in these HECs.

Then we estimate the Debye temperature ($\Theta_D$) through the equation $\Theta_D = (12\pi^4 nR/5\beta)^{1/3}$, where R has a value of 8.31 J/mol/K as the gas constant and n is the number of atoms per formula unit (n = 1 + $x$ for $Ti_{0.2}Zr_{0.2}Nb_{0.2}Mo_{0.2}Ta_{0.2}C_x$ samples). It yields $\Theta_D$ = 715 K and 647 K for $x$ = 1 and 0.8, respectively. With the $T_c$ and $\Theta_D$, we can obtain the electron-phonon coupling constant ($\lambda_{ep}$) through the McMillan formula, $\lambda_{ep} = \frac{1.04 + \mu^* \ln\left(\frac{\Theta_D}{1.45 T_c}\right)}{(1 - 0.62\mu^*) \ln\left(\frac{\Theta_D}{1.45 T_c}\right) - 1.04}$, where $\mu^*$ represents the Coulomb pseudopotential parameter and is typically given a value of 0.13.[46,47] Based on the obtained values, the superconducting parameter $\lambda_{ep}$ = 0.49 for $x$ = 1 and $\lambda_{ep}$ = 0.46 for $x$ = 0.8. The $\lambda_{ep}$ values suggest that $Ti_{0.2}Zr_{0.2}Nb_{0.2}Mo_{0.2}Ta_{0.2}C_x$ HECs are weak-coupling superconductors. In crystalline materials, electron-phonon coupling is a ubiquitous many-body interaction that drives conventional superconductivity. The phonon mechanism is responsible for the electron-electron coupling and, hence, the cause of superconductivity.[48-50] In the $Ti_{0.2}Zr_{0.2}Nb_{0.2}Mo_{0.2}Ta_{0.2}C_x$ system, the electron-phonon coupling strength weakens as the $T_c$ decreases.

To further investigate the superconductivity of HEC, we performed the high-pressure resistance measurements for $Ti_{0.2}Zr_{0.2}Nb_{0.2}Mo_{0.2}Ta_{0.2}C$ HEC. **Figure 5**a shows the typical resistance curves of $Ti_{0.2}Zr_{0.2}Nb_{0.2}Mo_{0.2}Ta_{0.2}C$ HEC under various pressures up to 82.5 GPa. It is seen that the superconducting transitions of $Ti_{0.2}Zr_{0.2}Nb_{0.2}Mo_{0.2}Ta_{0.2}C$ HEC subjected to different pressures are sharp, and the zero-resistance state remains present throughout the full range of pressures applied (see **Figure 5**b). The $T_c$ shows only a slight change from its ambient-pressure value of 4.15 K to 3.95 K at 82.5 GPa. The pressure-dependent $T_c$ for $Ti_{0.2}Zr_{0.2}Nb_{0.2}Mo_{0.2}Ta_{0.2}C$ HEC is mapped in the phase diagram in Figure 5c. We see a robust SC against high physical pressure in $Ti_{0.2}Zr_{0.2}Nb_{0.2}Mo_{0.2}Ta_{0.2}C$ HEC, with a slight $T_c$ variation of 0.3 K within 82.5 GPa. A similar phenomenon was also observed in $(TaNb)_{0.67}(HfZrTi)_{0.33}$ HEA.[51] This makes superconducting HECs also promising candidates for new applications under extreme conditions.

The lattice parameter of the $Ti_{0.2}Zr_{0.2}Nb_{0.2}Mo_{0.2}Ta_{0.2}C$ fitted by the Birch-Murnaghan equation of state is 4.484 Å, which is in consistent with the experimental lattice parameter (a = 4.4573(3) Å) (see Figure S4). For simplicity, the lattice parameters of $Ti_{0.2}Zr_{0.2}Nb_{0.2}Mo_{0.2}Ta_{0.2}C_x$ are fixed to the experimentally refined lattice constants. The total density of states (TDOS), the local density of states (DOS), and the partial DOS for $Ti_{0.2}Zr_{0.2}Nb_{0.2}Mo_{0.2}Ta_{0.2}C_x$ are shown in **Figure 6**. In this work, the supercell contains 64 atoms, so that the experimental doping ratio $r_{Ti} = r_{Zr} = r_{Nb} = r_{Mo} = r_{Ta} = 0.2$ cannot be obtained. We consider four different atomic arrangements' structural configurations (the doping ratio is equal to 0.1875 for three elements and 0.21875 for the remaining two.) for investigating the influence of the disorder on the electronic properties of the $Ti_{0.2}Zr_{0.2}Nb_{0.2}Mo_{0.2}Ta_{0.2}C_x$. The overall shape of the averaged TDOS for $x = 1$ and $x = 0.8$ are pretty similar, while is quite different near the Fermi level. The TDOS passing through the Fermi level suggests its typical metallic properties (see **Figure 6**a and **Figure 6**b). The local DOS diagram shows that the Ti, Zr, Nb, Mo, and Ta atoms are the most significant contribution to TDOS near the Fermi level. In contrast, the contribution from the C atoms is relatively modest. The d orbital of M and p orbital of C electrons are highly hybridized below the Fermi level. As displayed in Figure 6c-f, the projected DOS with angular momentum reveals that the *d*-electrons of M elements are the main contributions, i.e., 3*d* for Ti, 4*d* for Zr, Nb, Mo, and 5*d* for Ta. These results indicate that the superconductivity may mainly originate from the *d*-electrons of Ti, Zr, Nb, Mo, and Ta.

**Figure 7** shows the electronic band structures of $Ti_{0.2}Zr_{0.2}Nb_{0.2}Mo_{0.2}Ta_{0.2}C$ ($x = 1$). We first study the band structures of $Ti_{0.2}Zr_{0.2}Nb_{0.2}Mo_{0.2}Ta_{0.2}C$ without SOC. The three different structures are displayed in Figure S5 of Supporting Information. As is shown in Figure 7a-c, there exist six linear band intersections along G–X, G–Y, and G–Z directions at about - 0.75 eV (Type -II Dirac-like points (DPs) are denoted by the black circle rectangles). As shown in Figure 7d-e, the linear band intersection along G–X, G–Y, and G–Z directions is not split by considering the SOC, while three linear band intersections along the G–X1, G–Y1, and G–Z1 directions are lightly split (denoted by green circles). The projected band structures of $Ti_{0.2}Zr_{0.2}Nb_{0.2}Mo_{0.2}Ta_{0.2}C$ show that the DPs mainly contribute from the *d* orbitals of transition metals M and the *p* orbitals of C. As shown in Figure 7h, the positions of the DPs are sensitive to the strain. Therefore, we propose that the $Ti_{0.2}Zr_{0.2}Nb_{0.2}Mo_{0.2}Ta_{0.2}C$ is a topological superconductor candidate. Compared with the case of $Ti_{0.2}Zr_{0.2}Nb_{0.2}Mo_{0.2}Ta_{0.2}C$, the trivial electronic band structures of $Ti_{0.2}Zr_{0.2}Nb_{0.2}Mo_{0.2}Ta_{0.2}C_{0.8}$ are shown in Figure S6.

## 4. Conclusion

In conclusion, we have reported synthesized and characterized new HEC superconductors $Ti_{0.2}Zr_{0.2}Nb_{0.2}Mo_{0.2}Ta_{0.2}C_x$. Both $Ti_{0.2}Zr_{0.2}Nb_{0.2}Mo_{0.2}Ta_{0.2}C$ ($x = 1$) and $Ti_{0.2}Zr_{0.2}Nb_{0.2}Mo_{0.2}Ta_{0.2}C_{0.8}$ ($x = 0.8$) are discovered to be bulk superconductors with $T_c$ values of 4.00 and 2.65 K, respectively. The derived superconducting parameters show that $Ti_{0.2}Zr_{0.2}Nb_{0.2}Mo_{0.2}Ta_{0.2}C_x$ are type-II BCS weak-coupling superconductors. We observed a robust SC against high physical pressure in $Ti_{0.2}Zr_{0.2}Nb_{0.2}Mo_{0.2}Ta_{0.2}C$ HEC, with a slight $T_c$ variation of 0.3 K within 82.5 GPa. The first-principles calculations show that the DPs exist in the electronic band structures of $Ti_{0.2}Zr_{0.2}Nb_{0.2}Mo_{0.2}Ta_{0.2}C$. And the DPs are mainly contributed by the *p* orbitals of C and the *d* orbitals of transition metals M. The research results not only expand the new physical properties of HECCs but also provide a new material platform for studying the coupling between SC and topological physics.

**Acknowledgements**

The authors acknowledge productive conversations with Kui Jin at the Institute of Physics Chinese Academy of Sciences. Xunwu Hu thanks Cui-Qun Chen and Zequan Wang for their helpful discussions. This work is supported by the National Natural Science Foundation of China (12274471, 11922415), Guangdong Basic and Applied Basic Research Foundation (2022A1515011168, 2019A1515011718), the Key Research & Development Program of Guangdong Province, China (2019B110209003). Mebrouka Boubeche is supported by the Foreign Young Talents Program of China (22KW041C211). D. X. Yao and X. Hu are supported by NKRDPC-2022YFA1402802, NKRDPC-2018YFA0306001, NSFC-11974432, NSFC-92165204, Leading Talent Program of Guangdong Special Projects (201626003), and Shenzhen International Quantum Academy (Grant No. SIQA202102). This work was supported by the Synergetic Extreme Condition User Facility (SECUF).

References

[1] N. B. Kopnin, M. M. Salomaa, *Phys. Rev. B* **1991**, 44, 9667.

[2] N. Read, D. Green, *Phys. Rev. B* **2000**, 61, 10267.

[3] J. R. Badge, Y. Quan, M. C. Staab, S. Sumita, A. Rossi, K. P. Devlin, K. Neubauer, D.S. Shulman, J. C. Fettinger, P. Klavins, S. M. Kauzlarich, D. Aoki, I. M. Vishik, W. E. Pickett, V. Taufour, *Commun. Phys*. **2022**, 5, 22.

[4] X.-L. Qi, S.-C. Zhang, *Rev. Mod. Phys*. **2011**, 83, 1057.

[5] A. Y. Kitaev, *Ann. Phys*. **2003**, 303, 2.
[6] J.-P. Xu, M.-X. Wang, Z. L. Liu, J.-F. Ge, X. Yang, C. Liu, Z. A. Xu, D. Guan, C. L. Gao, D. Qian, Y. Liu, Q.-H. Wang, F.-C. Zhang, Q.-K. Xue, J.-F. Jia, *Phys. Rev. Lett*. **2015**, 114, 017001.
[7] M.-X. Wang, C. Liu, J.-P. Xu, F. Yang, L. Miao, M.Y. Yao, C, L, Gao, C. Shen, X. Ma, X. Chen, Z.-A. Xu, Y. Liu, S.-C. Zhang, D. Qiao, J.-F. Jia, Q.-K. Xue, *Science* **2012**, 336, 52.
[8] Y. S. Hor, A. J. Williams, J. G. Checkelsky, P. Roushan, J. Seo, Q. Xu, H. W. Zandbergen, A. Yazdani, N. P. Ong, R. J. Cava, *Phys. Rev. Lett*. **2010**, 104, 057001.
[9] P. P. Kong, J. L. Zhang, S. J. Zhang, J. Zhu, Q. Q. Liu, R. C. Yu, Z. Fang, C. Q. Jin, W. G. Yang, X. H. Yu, J. L. Zhu, Y. S. Zhao, *J. Phys. Condens. Matter*. **2013**, 25, 3662204.
[10] X. Zhang, K.-H. Jin, J. Mao, M. Zhao, Z. Liu, F. Liu, npj Comput. Mater. **2021**, 7, 44.
[11] L. A.Wray, S.-Y. Xu, Y. Xia, Y. S. Hor, D. Qian, A. V. Fedorov, H. Lin, A. Bansil, R. J. Cava, M. Z. Hasan, *Nat. Phys*. **2010**, 6, 855.
[12] G. Du, J. Shao, X. Yang, Z. Du, D. Fang, J. Wang, K. Ran, J. Wen, C. Zhang, H. Yang, Y. Zhang, H.-H. Wen, *Nat. Commun*. **2017**, 8, 14466.
[13] Z. Liu, X. Yao, J. Shao, M. Zuo, L. Pi, S. Tan, C. Zhang, Y. Zhang, *J. Am. Chem. Soc*. **2015**, 137, 10512.
[14] S. Sasaki, Z. Ren, A. A. Taskin, K. Segawa, L. Fu, Y. Ando, *Phys. Rev. Lett*. **2012**, 109, 217004.
[15] P. Hosur, X. Dai, Z. Fang, X.-L. Qi, *Phys. Rev. B* **2014**, 90, 045130.
[16] Z. Chi, X. Chen, C. An, L. Yang, J. Zhao, Z. Feng, Y. Zhou, Y. Zhou, C. Gu, B. Zhang, Y. Yuan, C. Kenney-Benson, W. Yang, G. Wu, X. Wan, Y. Shi, X. Yang, dan Z. Yang, *npj Quantum Mater*. **2018**, 3, 28.
[17] S. Zhu, L. Kong, L. Cao, H. Chen, M. Papaj, S. Du, Y. Xing, W. Liu, D. Wang, C. Shen, F. Yang, J. Schneeloch, R. Zhong, G. Gu, L. Fu, Y.-Y. Zhang, H. Ding, H.-J. Gao, *Science* **2020**, 367, 189.

[18]C.-K. Chiu, T. Machid, Y. Huang, T. Hanaguri, F.-C, Zhang, *Sci. Adv*. **2020**, 6, eaay0443.

[19]G. Xu, B. Lian, P. Tang, X.-L. Qi, S.-C. Zhang, *Phys. Rev. Lett*. **2016**, 117, 047001.

[20]W. Liu, L. Cao, S. Zhu, L. Kong, G. Wang, M. Papaj, P. Zhang, Y.-B. Liu, H. Chen, G. Li, F. Yang, T. Kondo, S. Du, G.-H. Cao, S. Shin, L. Fu, Z. Yin, H.-J. Gao, H. Ding, *Nat. Commun*. **2020**, 11, 5688.

[21]L. Kong, L. Cao, S. Zhu, M. Papaj, G. Dai, G. Li, P. Fan, W. Liu, F. Yang, X. Wang, S. Du, C. Jin, L. Fu, H.-J. Gao, H. Ding, *Nat. Commun*. **2021**, 21, 4146.

[22]M. Li, G. Li, L. Cao, X. Zhou, X. Wang, C. Jin, C.-K. Chiu, S. J. Pennycook, Z. Wang, H.-J. Gao, *Nature* **2022**, 606, 890.

[23]D. Shen, C. N. Kuo, T. W. Yang, I. N. Chen, C. S. Lue, L. M. Wang, *Commun. Mater*. **2020**, 1, 56.

[24]N. K. Karn, M. M. Sharma, V. P. S. Awana, *Supercond. Sci. Tech*. **2022**, 35, 114002.

[25]F. Martín-Vega, E. Herrera, B. Wu, V. Barrena, F. Mompeán, M. García-Hernández, P. C. Canfield, A. M. Black-Schaffer, J. J. Baldoví, I. Guillamón, He. Suderow, *Phys. Rev. Research* **2022**, 4, 023241.

[26]C. Q. Xu, B. Li, L. Zhang, J. Pollanen, X. L. Yi, X. Z. Xing, Y. Liu, J. H. Wang, Z. Zhu, Z. X. Shi, Xiaofeng Xu, X. Ke, *Phys. Rev. B* **2021**, 104, 125127.

[27]M. Kim, C.-Z. Wang, K.-M. Ho, *Phys. Rev. B* **2019**, 99, 224506.

[28]L. Han, X. Shi, J. Jiao, Z. Yu, X. Wang, N. Yu, Z. Zou, J. Ma, W. Zhao, W. Xia, Y. Guo, *Chinese Phys. Lett*. **2022**, 39, 067101.

[29]T. Shang, J. Zhao, L.-H. Hu, J. Ma, D. J. Gawryluk, X. Zhu, H. Zhang, Z. Zhen, B. Yu, Q. Zhan, E. Pomjakushina, M. Shi, T. Shiroka, *Sci. Adv*. **2022**, 8, eabq6589.

[30]T. Shang, J. Z. Zhao, D. J. Gawryluk, M. Shi, M. Medarde, E. Pomjakushina, T. Shiroka, *Phys. Rev. B* **2020**, 101, 214518.

[31]D. Y. Yan, M. Yang, C, X. Wang, P. B. Song, C. J. Yi, Y. G. Shi, *Supercond. Sci. Tech*. **2021**, 34, 035025.

[32]Z. Cui, Y. Qian, W. Zhang, H. Weng, Z. Fang, *Chinese Phys. Lett*. **2020**, 37, 087103.

[33] A. Huang, A. D. Smith, M. Schwinn, Q. Lu, T.-R. Chang, W. Xie, H.-T. Jeng, G. Bian, *Phys. Rev. Mater*. **2018**, 2, 054205.
[34] D. Yan, D. Geng, Q. Gao, Z. Cui, C. Yi, Y. Feng, C. Song, H. Luo, M. Yang, M. Arita, S. Kumar, E. F. Schwier, K. Shimada, L. Zhao, K. Wu, H. Weng, L. Chen, X. J. Zhou, Z. Wang, Y. Shi, B. Feng, *Phys. Rev. B* **2020**, 102, 205117.
[35] J.-W. Yeh, S.-K. Chen, S.-J. Lin, J.-Y. Gan, T.-S. Chin, T.-T. Shun, C.-H. Tsau, S.-Y. Chang, *Adv. Eng. Mater*. **2004**, 6, 299-303.
[36] S. Akrami, P. Edalati, M. Fuji, K. Edalati, *Mater. Sci. Eng. R: Rep*. **2021**, 146, 100644.
[37] H. Xiang, Y. Xing, F.-Z. Dai, H. Wang, L. Su, L. Miao, G. Zhang, Y. Wang, X. Qi, L. Yao, H. Wang, B. Zhao, J. Li Y. Zhou, *J. Adv. Ceram*. **2021**, 10, 385.
[38] S.-C. Luo, W.-M. Guo, K. Plucknett, H.-T. Lin, *J. Am. Ceram. Soc*. **2022**, 11, 805.
[39] X. Yan, L. Constantin, Y. Lu, J.-F. Silvain, M. Nastasi, B. Cui, *J. Am. Ceram. Soc*. **2018**, 101, 4486.
[40] P. Sarker, T. Harrington, C. Toher, C. Oses, M. Samiee, J.-P. Maria, D. W. Brenner, K. S. Vecchio S. Curtarolo, *Nat. Commun*. **2018**, 9, 4980.
[41] F. von Rohr, M. J. Winiarski, J. Tao, T. Klimczuk, R. J. Cava, *Proc. Natl. Acad. Sci. U.S.A*. **2016**, 113, E7144.
[42] P. Koželj, S. Vrtnik, A. Jelen, S. Jazbec, Z. Jagličić, S. Maiti, M. Feuerbacher, W. Steurer, J. Dolinšek, *Phys. Rev. Lett*. **2014**, 113, 107001.
[43] R. Prozorov, V. G. Kogan, *Phys. Rev. Applied* **2018**, 10, 014030.
[44] L. Zeng, Z. Wang, J. Song, G. Lin, R. Guo, S.-C. Luo, S. Guo, K. Li, P. Yu, C. Zhang, W.-M. Guo, J. Ma, Y. Hou, H. Luo, *Adv. Funct. Mater*. **2023**, 2301929.
[45] H. Shu, W. Zhong, J. Feng, H. Zhao, F. Hong, B. Yue, (Preprint) arXiv:2307.16438, v1, submitted: Jul 2023.
[46] L. Zeng, X. Hu, S. Guo, G. Lin, J. Song, K. Li, Y. He, Y. Huang, C. Zhang, P. Yu, J. Ma, D.-X. Yao, H. Luo, *Phys. Rev. B* **2022**, 106, 134501.
[47] L. Zeng, X. Hu, N. Wang, J. Sun, P. Yang, M. Bouheche, S. Luo, Y. He, J. Cheng, D.-X. Yao, H. Luo, *J. Phys. Chem. Lett*. **2022**, 13, 2442.
[48] P. K. Jha, S. P. Sanyal, *Physica C* **1996**, 271, 6.

[49] P. K. Jha, S. P. Sanyal, *Physica C* **1996**, 261, 259.

[50] M. Talati, P. K. Jha, *Phys. Rev. B* **2006**, 74, 134406.

[51] J. Guo, H. Wang, F. von Rohr, Z. Wang, S. Cai, Y. Zhou, K. Yang, A. Li, S. Jiang, Q. Wu, R. J. Cava, L. Sun, *Proc. Natl. Acad. Sci. U.S.A.* **2017**, 114, 13144.

[52] S.-C. Luo, W.-M. Guo, Z.-L. Fang, K. Plucknett, H.-T. Lin, *J. Eur. Ceram. Soc.* **2022**, 42, 336.

[53] H. K. Mao, J. Xu, P. M. Bell, *J. Geophys. Res.* **1986**, 91, 4673.

[54] A. van de Walle, *Calphad* **2009**, 33, 266.

[55] G. Kresse, J. Hafner, *Phys. Rev. B* **1993**, 47, 558.

[56] G. Kresse, J. Furthmüller, *Phys. Rev. B* **1996**, 54, 11169.

[57] J. P. Perdew, K. Burke, M. Ernzerhof, *Phys. Rev. Lett.* **1997**, 78, 1396.

[58] P. E. Blöchl, *Phys. Rev. B* **1994**, 50, 17953.

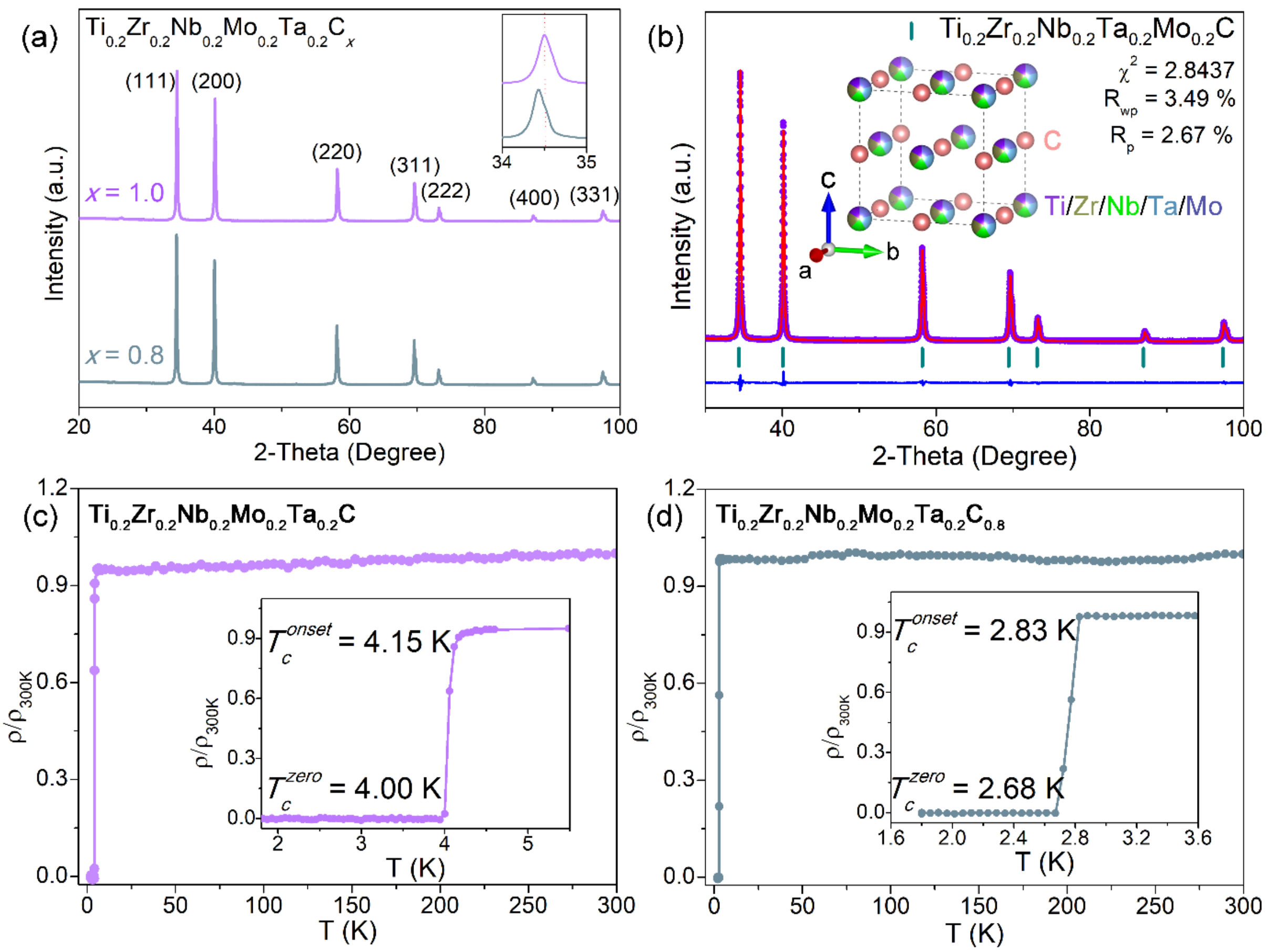

**Figure 1.** (a) PXRD patterns for $Ti_{0.2}Zr_{0.2}Nb_{0.2}Mo_{0.2}Ta_{0.2}C_x$ ($x$ = 1 and 0.8) samples. The inset shows the (111) reflections. (b) Rietveld refinement profile for $Ti_{0.2}Zr_{0.2}Nb_{0.2}Mo_{0.2}Ta_{0.2}C_x$ ($x$ = 1 and 0.8) samples. The inset displays the crystal structure. Electrical resistivity as functions of temperature for (c) $Ti_{0.2}Zr_{0.2}Nb_{0.2}Mo_{0.2}Ta_{0.2}C$ and (d) $Ti_{0.2}Zr_{0.2}Nb_{0.2}Mo_{0.2}Ta_{0.2}C_{0.8}$.

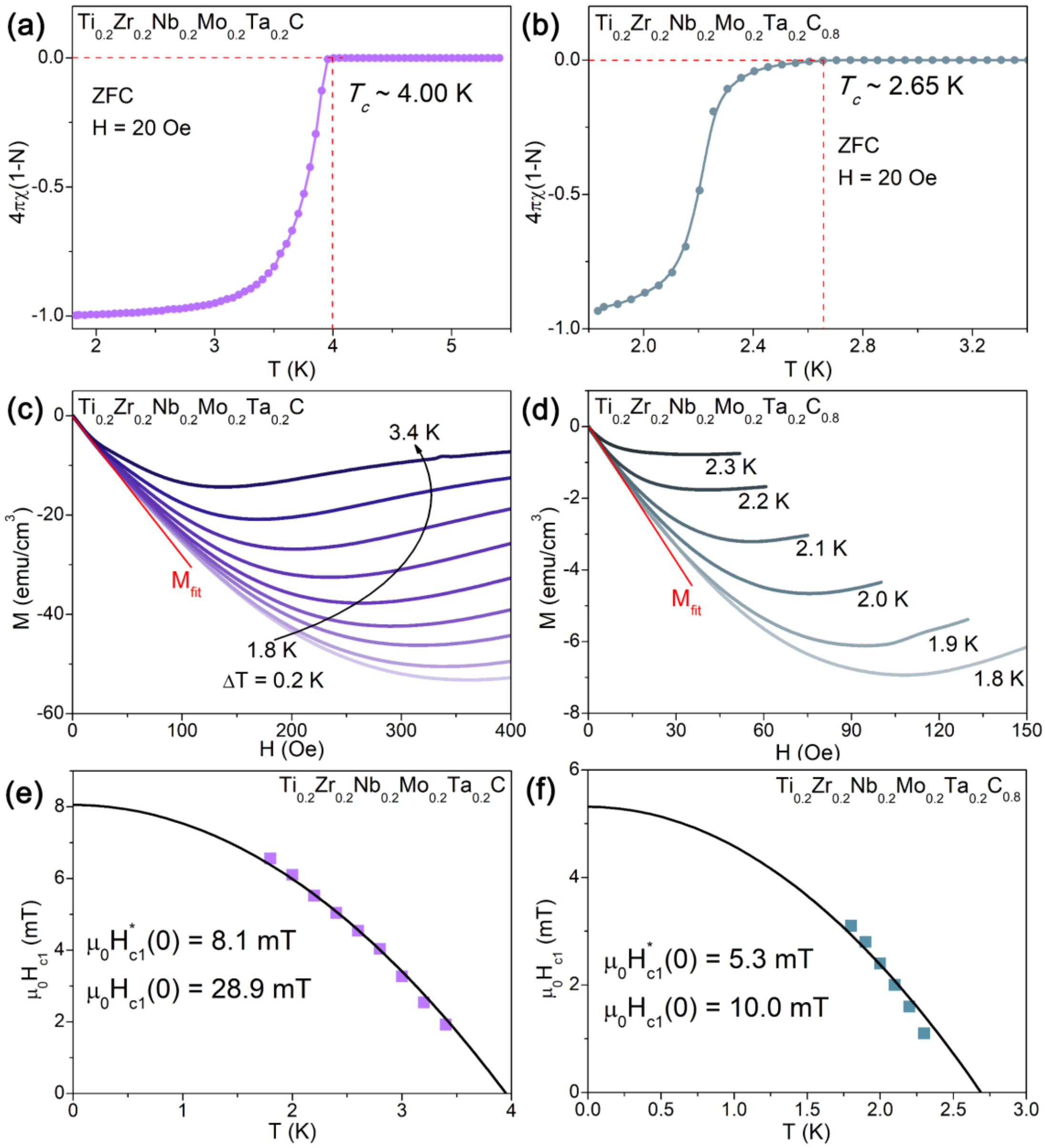

**Figure 2.** Temperature dependences of the zero-field-cooled (ZFC) volume magnetic susceptibility measured in a magnetic field of 20 Oe for (a) $Ti_{0.2}Zr_{0.2}Nb_{0.2}Mo_{0.2}Ta_{0.2}C$ and (b) $Ti_{0.2}Zr_{0.2}Nb_{0.2}Mo_{0.2}Ta_{0.2}C_{0.8}$. The field-dependent magnetization curves for (c) $Ti_{0.2}Zr_{0.2}Nb_{0.2}Mo_{0.2}Ta_{0.2}C$ and (d) $Ti_{0.2}Zr_{0.2}Nb_{0.2}Mo_{0.2}Ta_{0.2}C_{0.8}$. The temperature-dependent lower critical fields for (e) $Ti_{0.2}Zr_{0.2}Nb_{0.2}Mo_{0.2}Ta_{0.2}C$ and (f) $Ti_{0.2}Zr_{0.2}Nb_{0.2}Mo_{0.2}Ta_{0.2}C_{0.8}$.

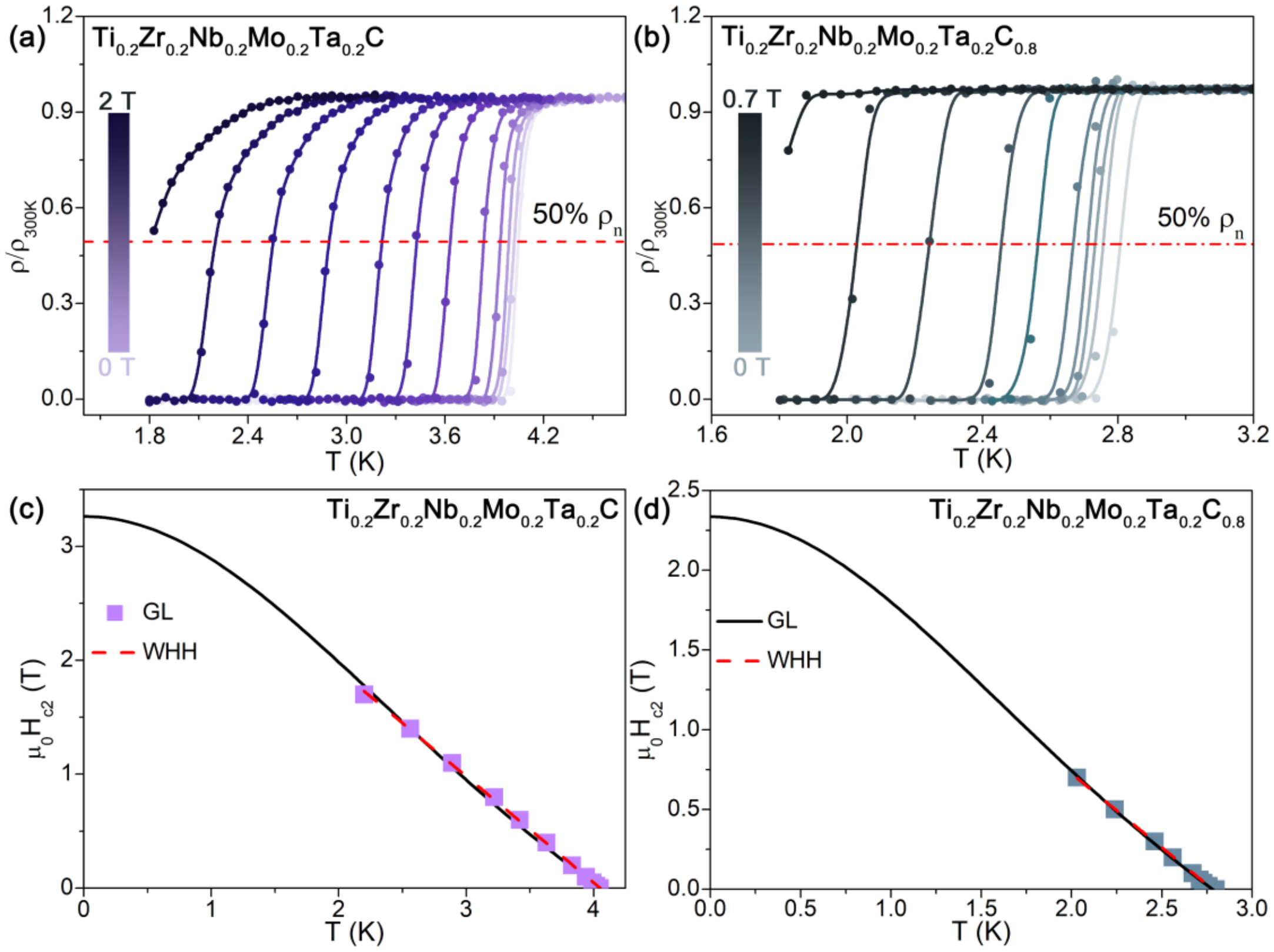

**Figure 3.** Low-temperature resistivity measurements in a variety of magnetic fields for (a) $Ti_{0.2}Zr_{0.2}Nb_{0.2}Mo_{0.2}Ta_{0.2}C$ and (b) $Ti_{0.2}Zr_{0.2}Nb_{0.2}Mo_{0.2}Ta_{0.2}C_{0.8}$. The temperature-dependent upper critical field for (c) $Ti_{0.2}Zr_{0.2}Nb_{0.2}Mo_{0.2}Ta_{0.2}C$ and (d) $Ti_{0.2}Zr_{0.2}Nb_{0.2}Mo_{0.2}Ta_{0.2}C_{0.8}$.

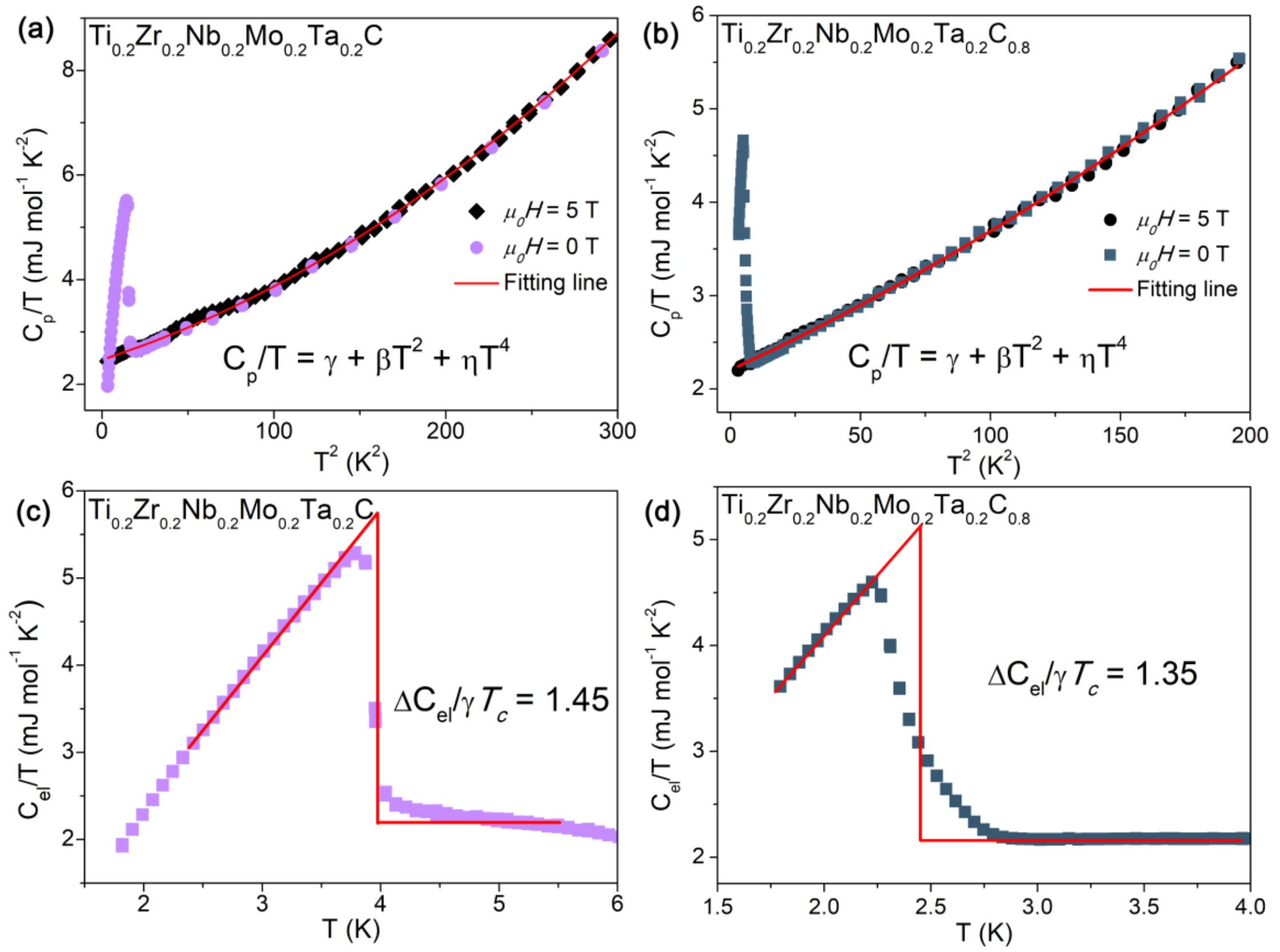

**Figure 4.** $C_p/T$ versus $T^2$ under 0 T and 5 T for (a) $Ti_{0.2}Zr_{0.2}Nb_{0.2}Mo_{0.2}Ta_{0.2}C$ and (b) $Ti_{0.2}Zr_{0.2}Nb_{0.2}Mo_{0.2}Ta_{0.2}C_{0.8}$. Temperature-dependent normalized electronic specific heats for (c) $Ti_{0.2}Zr_{0.2}Nb_{0.2}Mo_{0.2}Ta_{0.2}C$ and (d) $Ti_{0.2}Zr_{0.2}Nb_{0.2}Mo_{0.2}Ta_{0.2}C_{0.8}$.

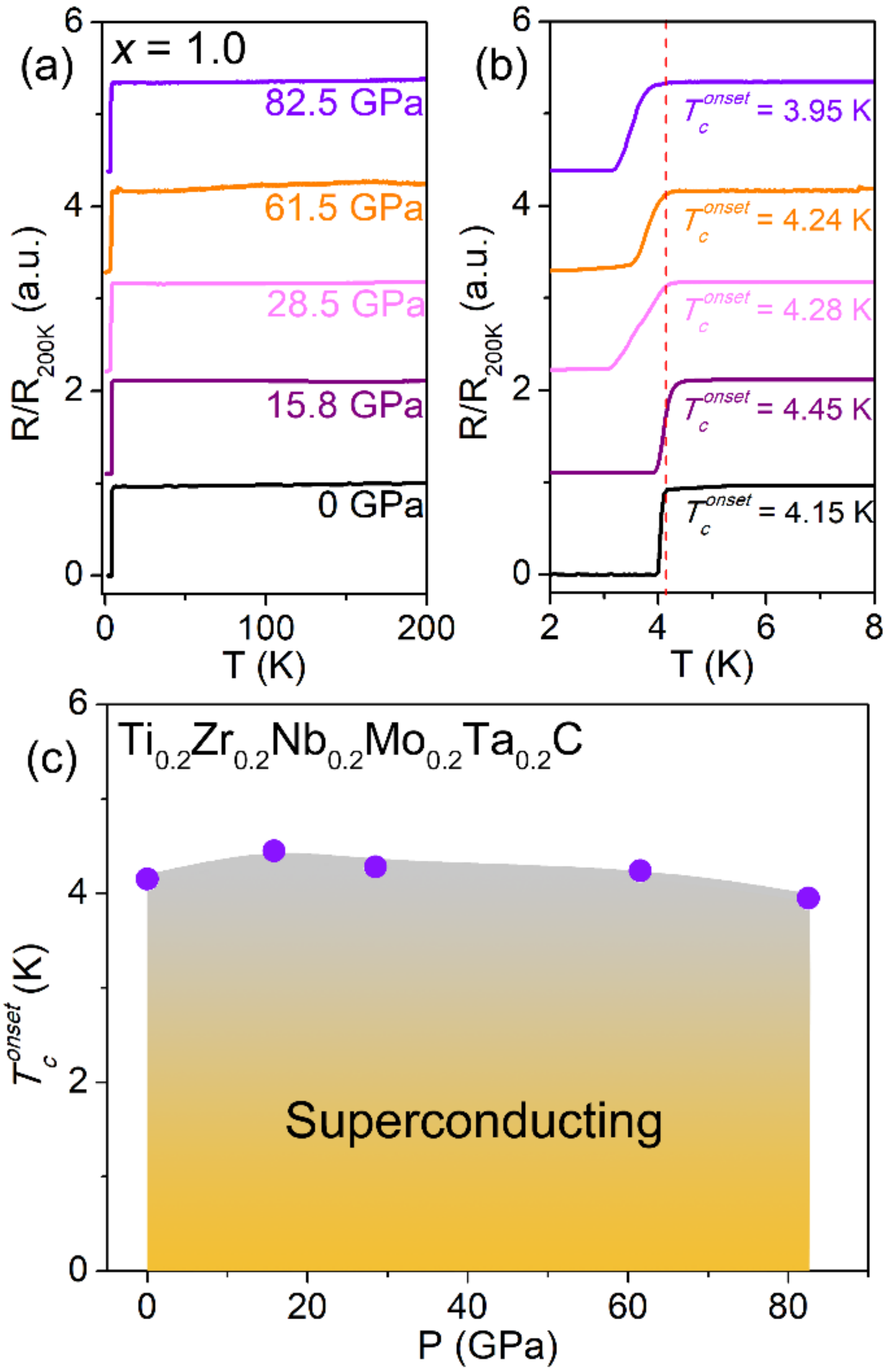

**Figure 5.** (a) Temperature dependence of resistance in $Ti_{0.2}Zr_{0.2}Nb_{0.2}Mo_{0.2}Ta_{0.2}C$ HEC in the pressure range of 0 - 82.5 GPa. (b) Temperature dependence of resistance in $Ti_{0.2}Zr_{0.2}Nb_{0.2}Mo_{0.2}Ta_{0.2}C$ HEC near the superconducting transition. (c) The phase diagram in $Ti_{0.2}Zr_{0.2}Nb_{0.2}Mo_{0.2}Ta_{0.2}C$ HEC is a function of pressure and temperature.

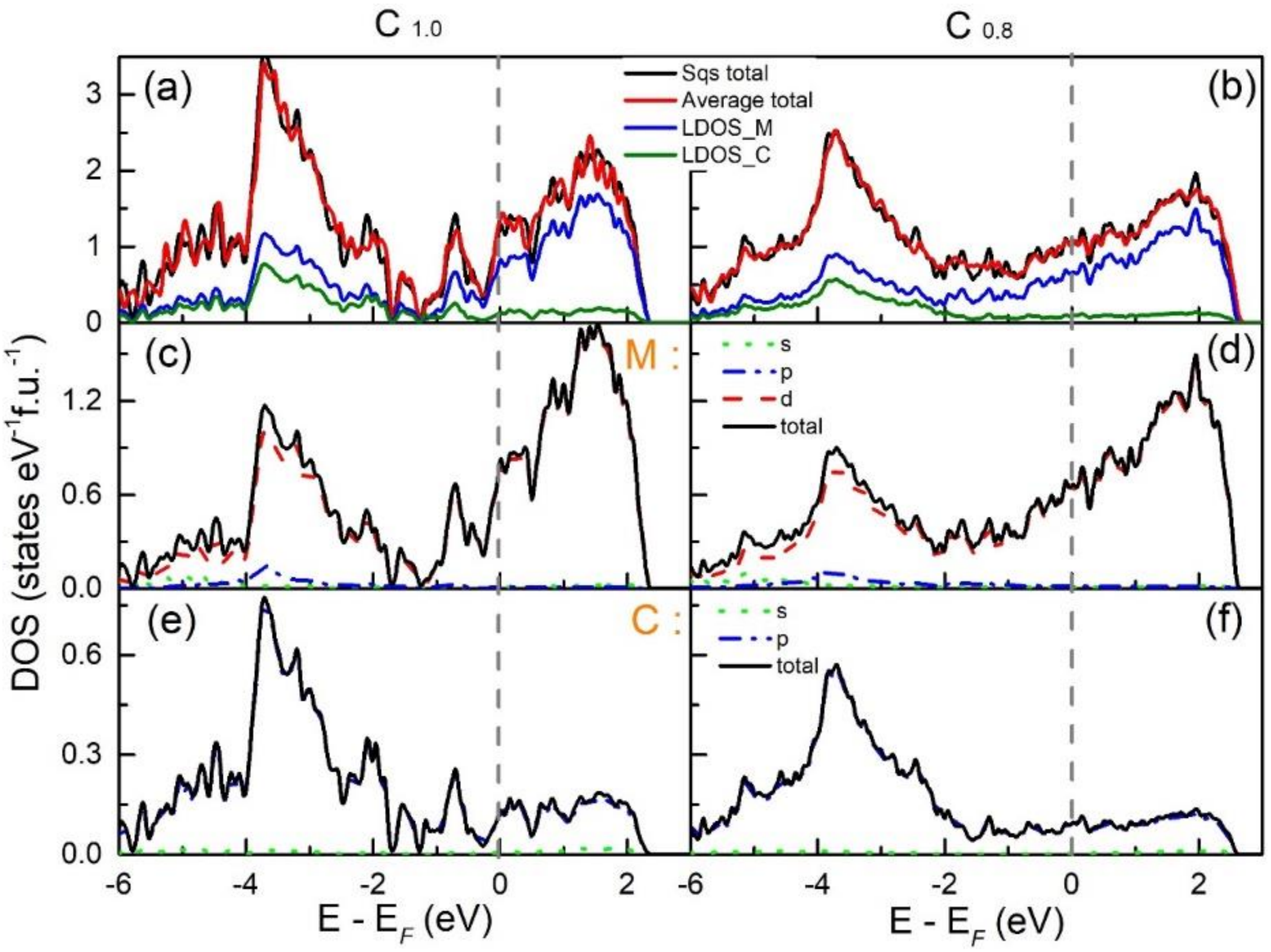

**Figure 6.** The Sqs and average DOS of $Ti_{0.2}Zr_{0.2}Nb_{0.2}Mo_{0.2}Ta_{0.2}C_x$ calculated by considering four assumed structures built by "mcsqs" code ($x$ = 1 for (a), (c), and (e), $x$ = 0.8 for (b), (d), and (f)). (a) - (b) Total and local DOS of $Ti_{0.2}Zr_{0.2}Nb_{0.2}Mo_{0.2}Ta_{0.2}C_x$. (c)-(f) Projected DOS with angular momentum decomposition of each element. The gray dashed lines indicate the Fermi level.

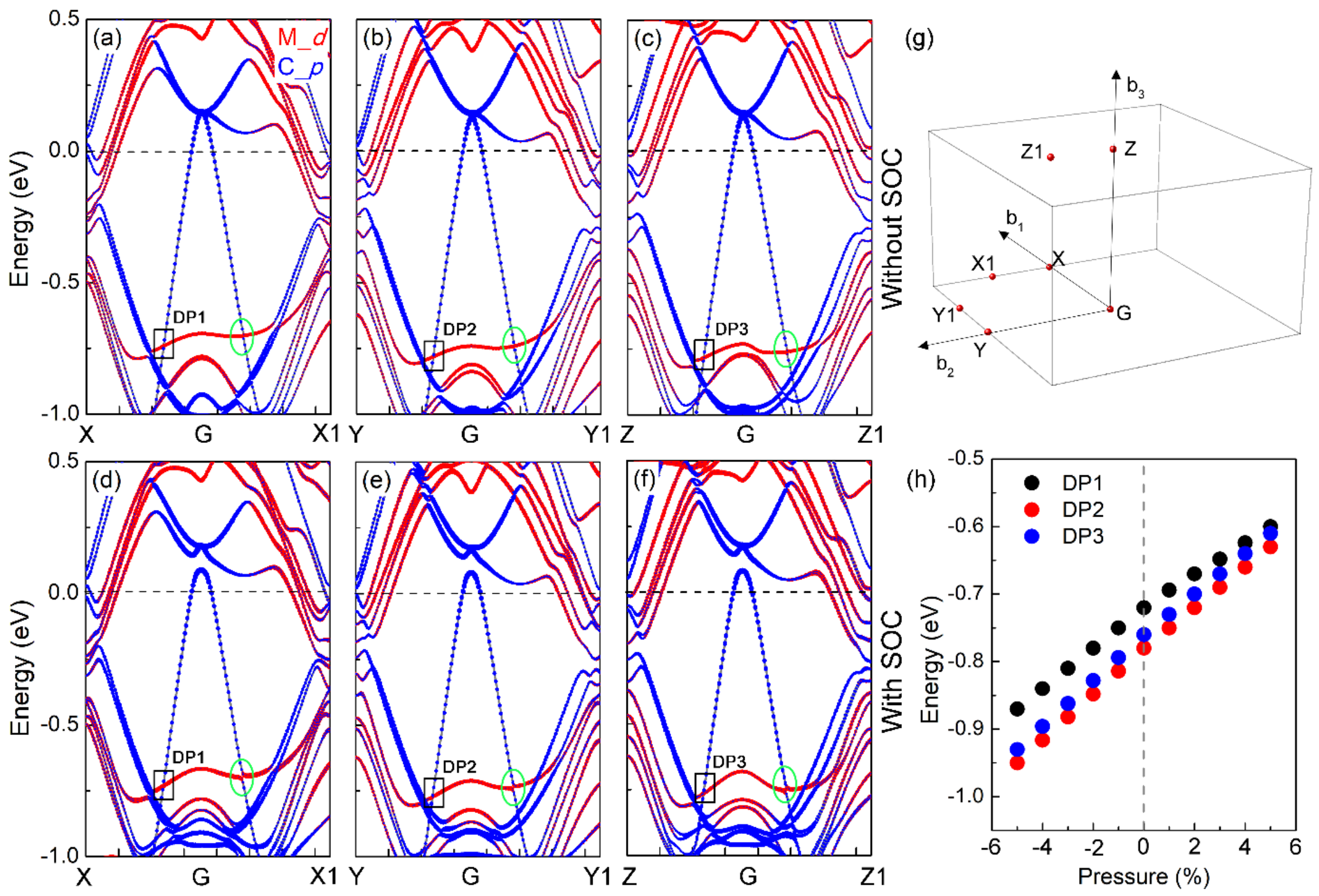

**Figure 7.** Electronic band structures of $Ti_{0.2}Zr_{0.2}Nb_{0.2}Mo_{0.2}Ta_{0.2}C$ ($x$ = 1) calculated by (a)(b)(c) ignoring and by (d)(e)(f) considering the spin-orbit coupling (SOC). The black rectangles indicate the type-II DPs. The three representative crystal structures are used in (a)-(c), shown in Figure S4. (g) The first Brillouin zone of $Ti_{0.2}Zr_{0.2}Nb_{0.2}Mo_{0.2}Ta_{0.2}C$. (h) Strain dependence of the relative energy at the position of type-II DPs. The Fermi level is indicated by the gray dashed lines. An amplified scaling factor of five is used for the C element in (a)-(f).

# Supporting Information

**Superconductivity in the high-entropy ceramics $Ti_{0.2}Zr_{0.2}Nb_{0.2}Mo_{0.2}Ta_{0.2}C_x$ with possible nontrivial band topology**

*Lingyong Zeng[1,#], Xunwu Hu[2,#], Yazhou Zhou[7,#], Mebrouka Boubeche[3], Ruixin Guo[4,5], Yang Liu[6], Si-Chun Luo[6], Shu Guo[4,5], Kuan Li[1], Peifeng Yu[1], Chao Zhang[1], Wei-Ming Guo[6], Liling Sun[7,*], Dao-Xin Yao[2,5*], Huixia Luo[1,*]*

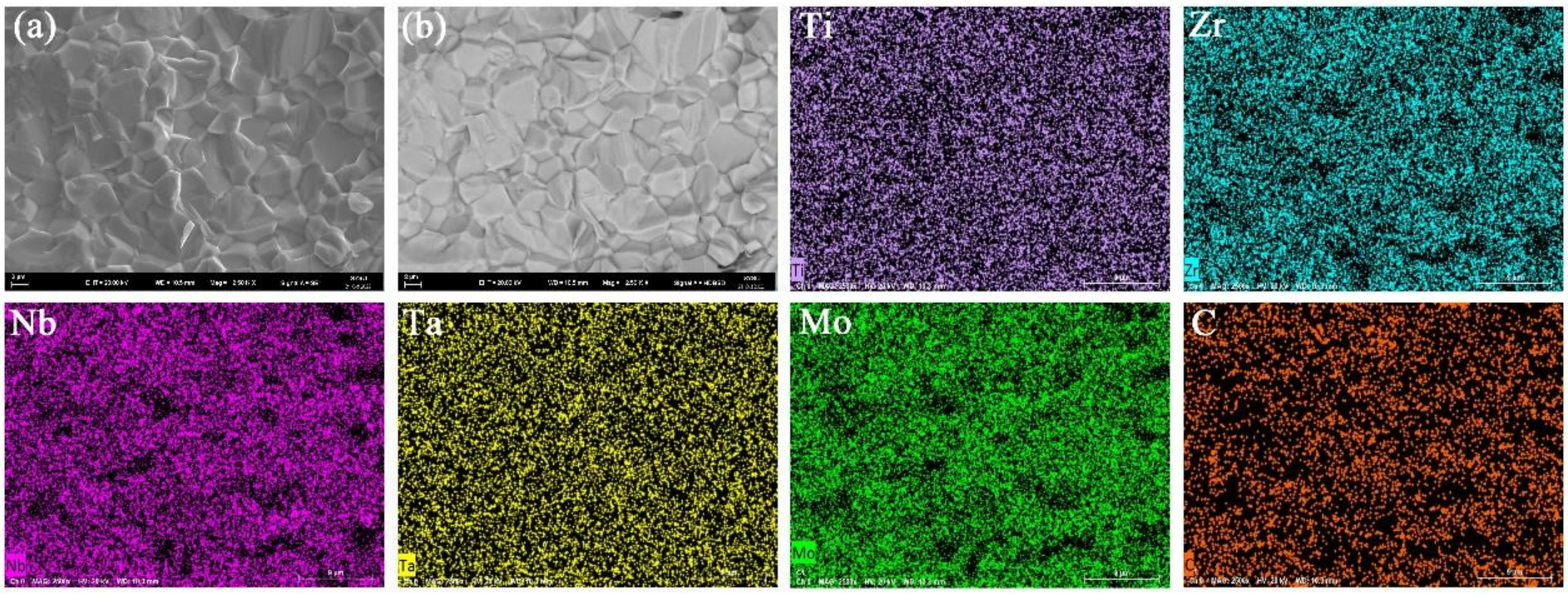

**Figure S1** (a) SEM, (b) BSEM images, and EDX mappings of $Ti_{0.2}Zr_{0.2}Nb_{0.2}Mo_{0.2}Ta_{0.2}C$.

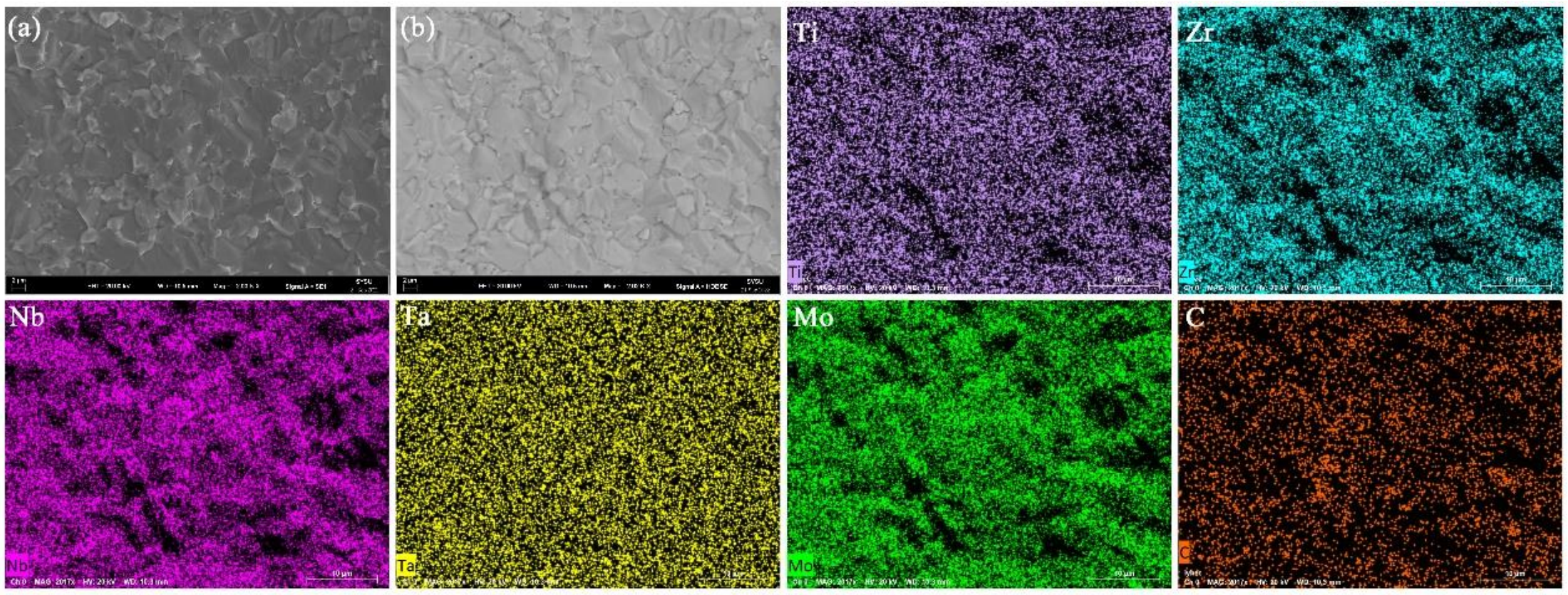

**Figure S2** (a) SEM, (b) BSEM images, and EDX mappings of $Ti_{0.2}Zr_{0.2}Nb_{0.2}Mo_{0.2}Ta_{0.2}C_{0.8}$.

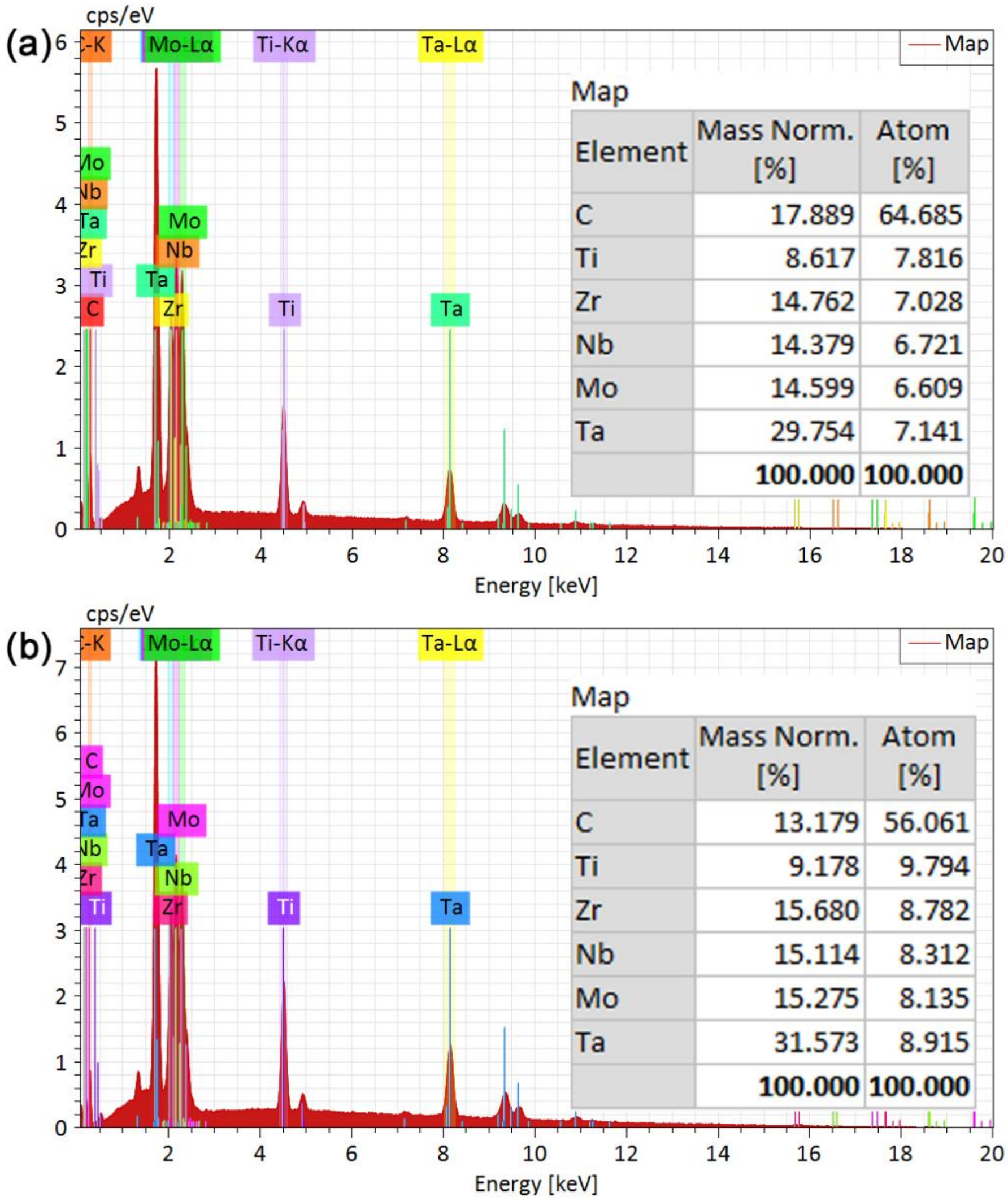

**Figure S3** EDX spectrum of (a) $Ti_{0.2}Zr_{0.2}Nb_{0.2}Mo_{0.2}Ta_{0.2}C$ and (b) $Ti_{0.2}Zr_{0.2}Nb_{0.2}Mo_{0.2}Ta_{0.2}C_{0.8}$.

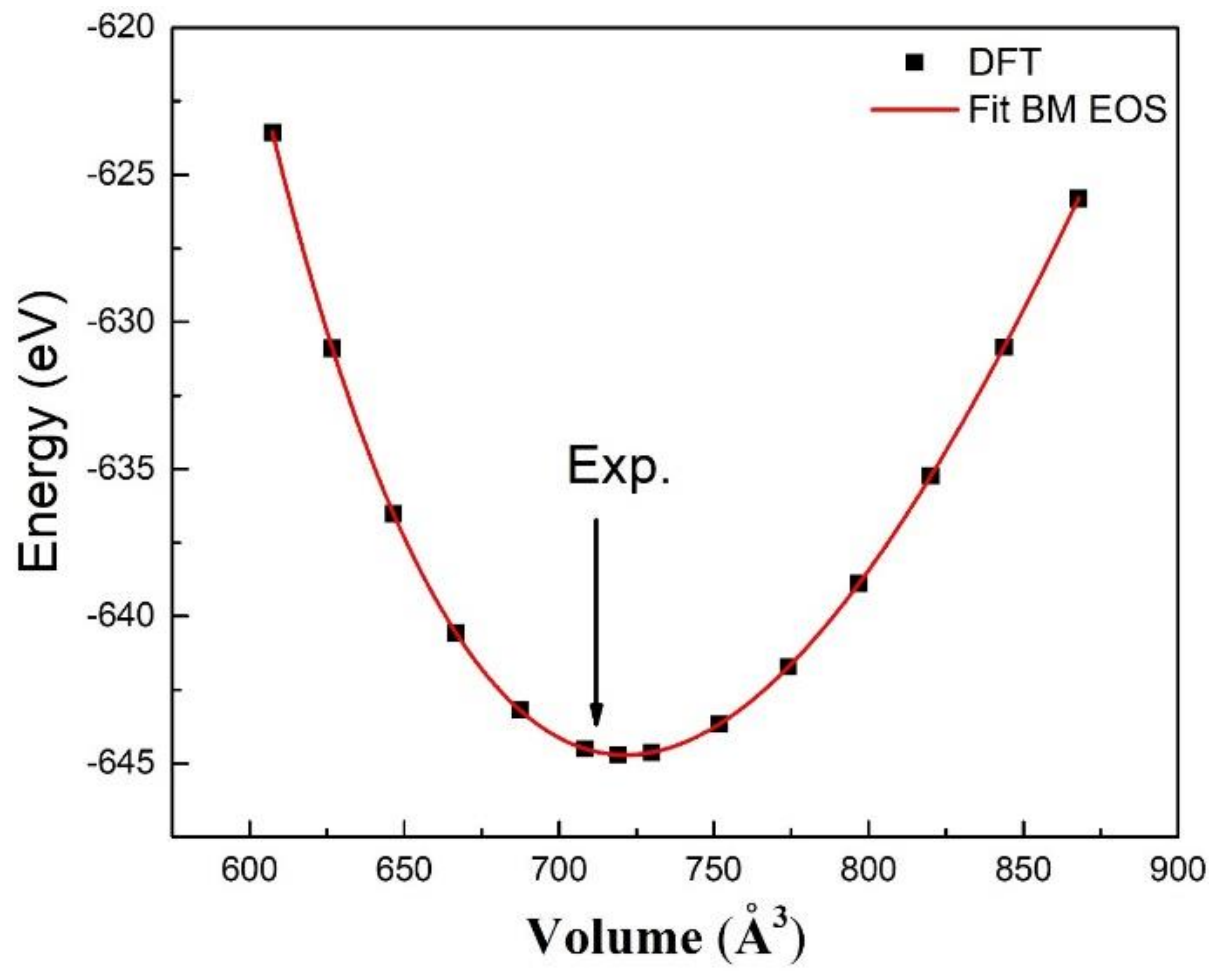

**Figure S4** The calculated total energy as a function of volume.

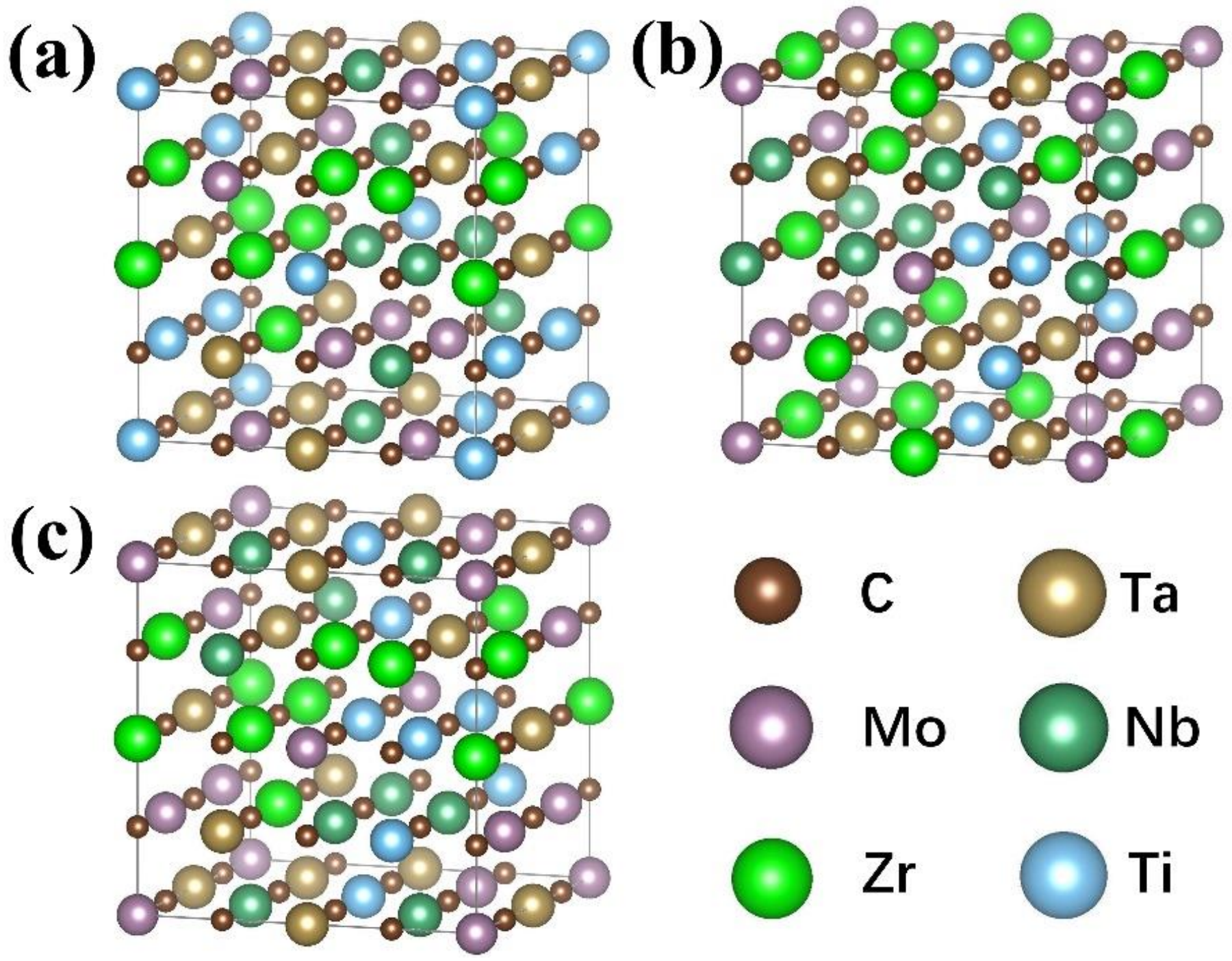

**Figure S5** (a)-(c) Crystal structures of the three representative structures for Fig. 7(a)-(c).

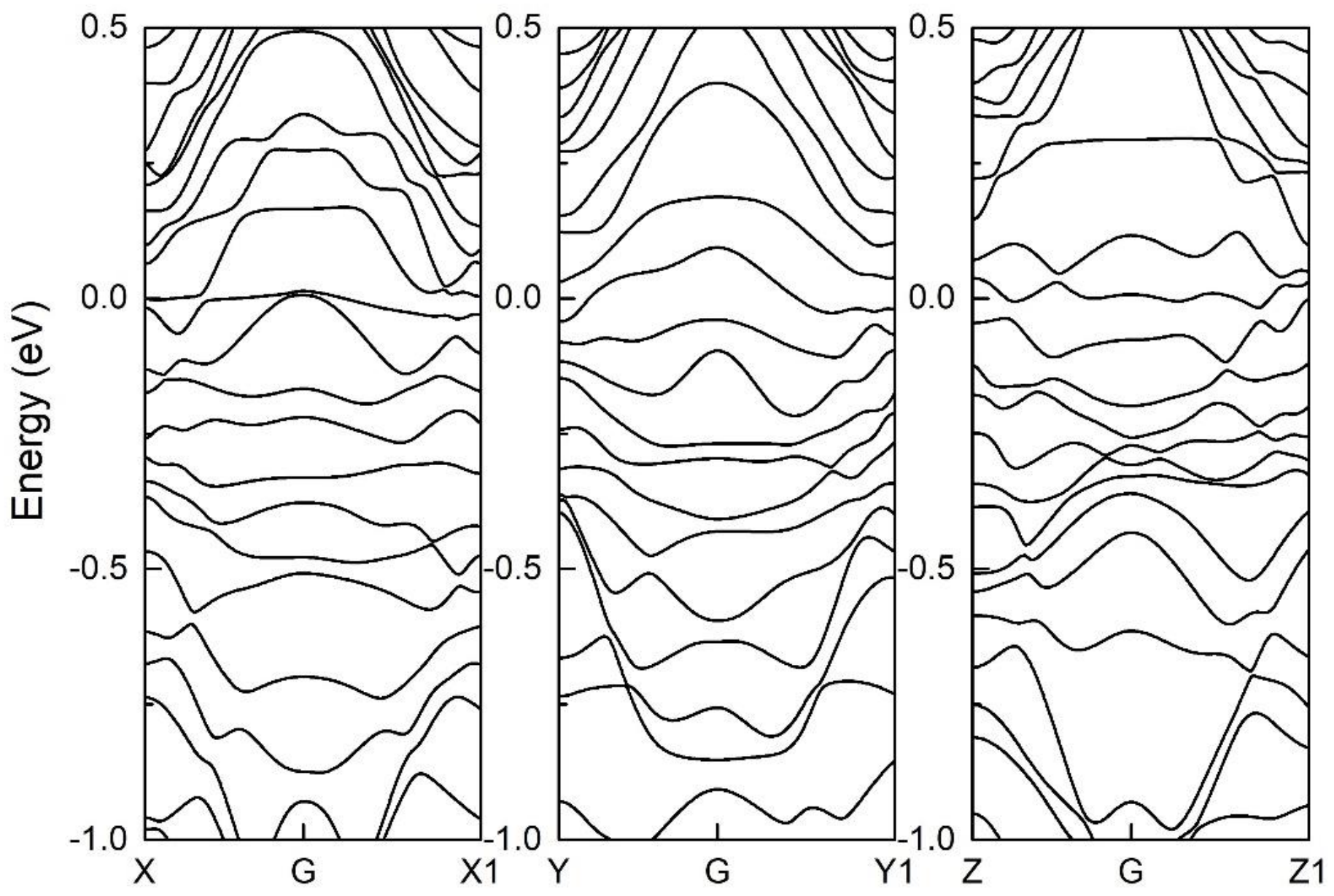

**Figure S6** Electronic band structures of $Ti_{0.2}Zr_{0.2}Nb_{0.2}Mo_{0.2}Ta_{0.2}C$ ($x$ = 0.8).

**Table S**1 The superconducting parameters of HECCs.

| Property | $Ti_{0.2}Zr_{0.2}Nb_{0.2}Mo_{0.2}Ta_{0.2}C$ | $Ti_{0.2}Zr_{0.2}Nb_{0.2}Mo_{0.2}Ta_{0.2}C_{0.8}$ | $Ti_{0.2}Zr_{0.2}Nb_{0.2}Hf_{0.2}Ta_{0.2}C$[43] | $Mo_{0.2}Nb_{0.2}Ta_{0.2}V_{0.2}W_{0.2}C_{0.9}$[44] | $Ta_{0.25}Ti_{0.25}Nb_{0.25}Zr_{0.25}C$[44] |
|---|---|---|---|---|---|
| $T_c$ (K) | 4.00 | 2.65 | 2.35 | 3.40 | 5.70 |
| $\mu_0H_{c1}$ (mT) | 28.9(6) | 10.0(2) | 26.1 | | |
| $\mu_0H_{c2}$ (T) | 3.2(6) | 2.3(4) | 0.5(1) | 3.4 | 1.5 |
| $\mu_0H^P$ (T) | 7.40 | 4.90 | 4.35 | 6.29 | 10.55 |
| $\xi_{GL}(0)$ (Å) | 100.5(3) | 118.6(5) | 261.8(4) | | |
| $\lambda_{GL}(0)$ (Å) | 1186 | 2192 | - | | |
| $\Theta_D$ (K) | 715 | 647 | 724 | | |
| $\lambda_{ep}$ | 0.49 | 0.46 | 0.54 | | |